\documentclass[12pt,reqno,oneside,titlepage]{amsart}

\newtheorem{theorem}{Theorem}[section]
\newtheorem{proposition}[theorem]{Proposition}
\newtheorem{definition}[theorem]{Definition}
\newtheorem{lemma}[theorem]{Lemma}
\newtheorem{corollary}[theorem]{Corollary}

\newcommand\lb{\left[}
\newcommand\rb{\right]}
\newcommand\lp{\left(}
\newcommand\rp{\right)}

\newcommand\cmp{\mathbf{p}}
\newcommand\cmpq{\mathbf{q}}

\newcommand\nodd{{\mathop{o}}}
\newcommand{\neven}{{\mathop{e}}}

\newcommand\Res{\mathop{\mathrm{Res}}}

\newcommand\E{\mathrm{e}}

\newcommand\sth{^{\text{th}}}

\newcommand\natnums{\mathbb{N}}

\newcommand\cH{\mathcal{H}}
\newcommand\cG{\mathcal{G}}
\newcommand\cJ{\mathcal{J}}
\newcommand\cT{\mathcal{T}}

\newcommand\tH{\widetilde{\cH}}

\newcommand\hT{\widehat{\cT}}
\newcommand\hH{\widehat{\cH}}

\newcommand\hP{\widehat{P}}
\newcommand\hQ{\widehat{Q}}

\newcommand\tphi{\tilde{\phi}}
\newcommand\hphi{\hat{\phi}}

\newcommand\tU{\tilde{U}}

\newcommand\tpsi{\tilde{\psi}}

\newcommand\trho{\tilde{\rho}}

\newcommand{\bisq}{b_{n,i}^{\, 2}}

\newcommand{\subover}[2]{\genfrac{}{}{0pt}{}{#1}{#2}}

\title[Spectral residues]{Spectral residues of second-order
  differential equations: a new methodology for summation identities
  and inversion formulas.}

\author{R. Milson\\Dept. Mathematics \&  Statistics\\Dalhousie University}


\address{Dept of Math., Dalhousie U., Halifax, Canada, B3H 3J5}

\email{milson@mathstat.dal.ca} 

\thanks{Email: {\tt milson@mathstat.dal.ca}}
\thanks{This research supported by a Dalhousie University startup grant.}

\begin{document}

\begin{abstract}
  The present article deals with differential equations with spectral
  parameter from the point of view of formal power series.  The
  treatment does not make use of the notion of eigenvalue, but
  introduces a new idea: the spectral residue.  
  
  The article focuses on second-order, self-adjoint problems.  In such
  a setting every potential function determines a sequence of spectral
  residues.  This correspondence is invertible, and gives rise to a
  combinatorial inversion formula.  Other interesting combinatorial
  consequences are obtained by considering spectral residues of
  exactly-solvable potentials of $1$-dimensional quantum mechanics.
  
  It is also shown that the Darboux transformation of $1$-dimensional
  potentials corresponds to a simple negation of the corresponding
  spectral residues.  This fact leads to another combinatorial
  inversion formula.  
  
  Finally, there is a brief discussion of applications.  The topics
  considered are enumeration problems and integrable systems.
\end{abstract}
\maketitle

\pagestyle{myheadings}

\section{Introduction}
There is a rich interplay between the methods and ideas of theoretical
physics and that of combinatorics.  The connection runs both ways:
combinatorial methods are applied to the study of physical models,
while ideas from physics are used to obtain results of a purely
combinatorial nature.

The present paper follows the second pattern.  The aim is to
demonstrate that certain methods for the solution of the
time-independent Scrh\"odinger equation lead naturally to interesting
identities and inversion formulas.  The idea that allows us to go from
the one domain to the other is that of a spectral residue.  The
application of differential operators and residue techniques to
combinatorial identities and inversion formulas is well documented
\cite{Riordan,Goulden,Huang,Olver}.  However, the notion of a spectral
residue seems to be unknown, and therefore deserves to be considered
in some detail.

The impetus for spectral residues comes from the following problem: to
give a formal series treatment of differential equations with spectral
parameter.  It is far from clear whether the notion of eigenvalue is
meaningful in a formal setting, and so we take a different approach.
Consider a linear differential operator $T$ in one independent
variable $z$, with spectral parameter $\nu$, and define $\phi(z;\nu)$
as the formal series solution of $T[\phi]=0.$ As the analogue of the
spectrum of $T$, let us take the set of values of $\nu$ for which some
of the coefficients of $\phi$ are singular.  This spectral set is not
of interest in itself; in the cases considered here it is just the set
of natural numbers.  More interesting are the residue of the
coefficients of the solution at the singular values of $\nu$; these we
will call the spectral residues of the operator.

Let us restrict our attention to classical quantum-mechanics and
consider second-order, differential equations that are amenable to the
usual method of undetermined coefficients (aka the method of
Frobenius):
$$T[\phi]=z^2 \phi''(z) + (1-\nu + P(z))z\phi'(z) + Q(z)\phi(z)=0,\;
\phi(0)=1.$$
Here $P$ and $Q$ are formal power series in $z$ with
vanishing constant term, and $\nu$ is the spectral parameter.  The
roots of the indicial equation are $0$ and $\nu$. Hence if $\nu$ is a
natural number, then, in general, the solution is singular.  In this
way there arises a sequence of spectral residues, one for each natural
number.  If we can solve the operator equation $T[\phi]=0$, then we
will be able to obtain information about this sequence.

From the very beginning of quantum mechanics, the study of
Schr\"o\-dinger's equation was marked by the search for exact solutions.
The two outstanding methods in this endeavor are first, the theory of
hypergeometric functions\cite{Bateman}, and second the Darboux
transformation\cite{Darboux,Cooper}.  The first method furnishes a
large class of exactly-solvable potentials whose eigenfunctions can be
given in terms of either the hypergeometric or the confluent
hypergeometric functions\cite{Flugge}.  All the classical
exactly-solvable potentials: the Harmonic oscillator, the Coulomb, the
Morse, the Eckart potentials, are of this type.

As for the second method, the essential feature of the Darboux
transformation is that it changes one potential function into another
in such a way that the eigenfunctions of the one can be explicitly
constructed from the eigenfunctions of the other.  By means of this
pairing exactly solvable potentials yield new exactly solvable
potentials.  Combining the two methods it becomes possible to describe
a very large class of exactly solvable potentials.

Our goal is to examine these two methodologies from a combinatorial
standpoint by using spectral residues.  In this context the following
questions are quite natural.
\begin{enumerate}
\item Does the spectral residue sequence completely determine the
  potential?  If yes, can this relationship be described explicitly?
  
\item What are the spectral residue sequences corresponding to
  potentials of hypergeometric type?  Does the exact-solvability of
  these potentials manifest itself at the level of spectral residues?
  
\item What is the effect of the Darboux transformation on the
  corresponding spectral residue sequence?
\end{enumerate}

We will see that all of these questions have satisfactory answers, and
that these answers manifest themselves as a novel class of summation
identities and inversion formulas.  The sums that occur in this
context are indexed by compositions, i.e. ordered partitions.  More
specifically, we will say that an $l$-part composition of a natural
number $n$ is an {\em ordered} list of $l$ positive integers
$\cmp=(p_1, p_2, \ldots, p_l)$ such that $$p_1+\ldots + p_l=n.$$
On
occasion we will write $|\cmp|$ instead of $n$, and $\ell(\cmp)$
instead of $l$.

Identities that feature sums indexed by  partitions
are not new \cite{Fine}\cite[e.g. \S 60]{Jordan}\cite{Moser}. All
of these references feature sums whose terms are symmetric functions
of the partition entries.  Thus, no generality is lost by working with
partitions rather than compositions. By contrast, the sums that arise in
connection with spectral residues have terms that are
functions of the partial sums of the composition entries;  the order
of the entries cannot be ignored. 

For a composition $\cmp=(p_1,\ldots,p_l)$ we define
$$s_\cmp=s_1 s_2 \ldots s_l,$$
where $s_j=p_1+\ldots+p_j$ is the
$j\sth$ left partial sum corresponding to $\cmp$.  Thus, if one
interprets $\cmp$ as a walk from $0$ to $n$ with steps
$p_1,\ldots,p_l$, then $s_\cmp$ is the product of numbers visited on
that walk.  Let $\cmp'$ denote the reversed composition
$(p_l,\ldots,p_2,p_1)$.  In this way $s_{\cmp'}$ is equal to the
product of visited numbers when one interprets $\cmp$ as a reverse
walk from $n$ down to $0$.

Let $U_1, U_2, U_3,\ldots $ be an infinite sequence of non-commuting
variables.  For $n=1,2,\ldots$ set
\begin{equation}
  \label{eq:rhodef}
  \rho_n = \sum_{|\cmp|=n} \frac{n}{s_\cmp\,s_{\cmp'}} \,U_\cmp
\end{equation}
where the index runs over all compositions $\cmp=(p_1,\ldots,p_l)$ of
$n$, and $U_\cmp$ is an abbreviation for the $l$-fold product $U_{p_1}
U_{p_2} \ldots U_{p_l}$.  Here are the first few elements of the
resulting sequence:
\begin{align*}
  \rho_1 &= U_1, \displaybreak[0]\\
  \rho_2 &= \frac12 U_2 + \frac12 U_1{}^2,\displaybreak[0]\\
  \rho_3 &= \frac13 U_3 + \frac16 (U_2 U_1 + U_1U_2)+\frac1{12}U_1{}^3,\displaybreak[0]\\
  \rho_4 &= \frac14 U_4 + \frac1{12} ( U_3 U_1 + U_1 U_3) + 
  \frac1{16} U_2{}^2 +\\
  & \quad + \frac1{48} ( U_1{}^2 U_2 + U_2 U_1{}^2) 
  +\frac{1}{36} U_1 U_2 U_1 + \frac1{144} U_1{}^4
\end{align*}
Evidently, equation \eqref{eq:rhodef} defines an invertible, nonlinear
mapping from the set of sequences $\{U_n\}$ to the set of sequences
$\{\rho_n\}$.
\begin{theorem}
  \label{th:rhoinv}
  The inverse mapping is given by
  \begin{equation}
    \label{eq:rhoinv}
    U_n = \sum_{|\cmp|=n}(-1)^{l-1}\, \frac{ n
    }{(p_1+p_2)(p_2+p_3)\ldots (p_{l-1}+p_l)}\,\rho_\cmp   
  \end{equation}
  where the
  sum is taken over all compositions $\cmp=(p_1,\ldots,p_l)$ of $n$,
    and $\rho_\cmp=\rho_{p_1}\rho_{p_2}\ldots \rho_{p_l}$.
\end{theorem}
What does this theorem have to do with potential functions and
spectral residues?  Letting $U(z)=U_1z+U_2z^2+\ldots$ and working with
the ``nearly'' self-adjoint equation
$$z^2\psi''(z) + (1-\nu) z\psi'(z) + U(z)\psi(z) = 0,$$
we will see
below that the corresponding spectral residues are given by equation
\eqref{eq:rhodef}.  The inversion formula given in Theorem
\ref{th:rhoinv} is motivated by question 1 above. The proof of the
formula will be given below in Section \ref{sect:specres}.  

Before continuing, we should remark that in classical quantum
mechanics the potential function $U(z)$ is scalar-valued, and hence
the coefficients $U_n$ commute.  Likewise the usual application of the
method of Frobenius is to series with commuting coefficients. However,
formula such as \eqref{eq:rhodef} and \eqref{eq:rhoinv} involve a sum
indexed by compositions, and as such are best treated in a
non-commutative setting.  Indeed, if the $U_n$ commuted, then the
natural form for the inversion formula would be a sum indexed by
partitions, i.e.  unordered lists of natural numbers, rather than
compositions.

The reader should therefore be warned that our discussion of spectral
residues will involve both the commutative and the non-commutative
settings.  In as much as commutativity is an extra assumption, we will
try to do as much as possible with non-commuting coefficients.
However, much of the discussion, in particular question 2 requires the
commutativity assumption.  As a consequence, we will find ourselves
jumping several times between the two different settings.

As regards question 2, we should first recall that the hypergeometric
equation has 3 parameters.  Making a suitable change of the parameter
variables one of them can be made into a spectral parameter.  Taking
spectral residues ``uses up'' this parameter, and as a result the
spectral residue sequence consists of polynomials in the remaining two
parameters.

We should also observe that the exact-solvability of the
hypergeometric potentials is really due to the fact that the
hypergeometric equation admits polynomial solutions for certain
discrete values of its parameters.  This observation has an important
consequence: it turns out that the spectral residues of hypergeometric
potentials factorize with respect to the two remaining parameters.
This in turn leads to interesting combinatorial identities.

Here is an example of this phenomenon associated to the potential
function
$$U(z) = \frac{vz}{(1-z)^2} = vz + 2vz^2+3vz^3+\ldots,$$
where the $v$
is a parameter. (N.B. The other hypergeometric parameter was
  dropped to simplify this example.  It will reappear in Section
  \ref{sect:espot}, where this sort of factorization is discussed
  fully.)  As per \eqref{eq:rhodef}, the spectral residues are given
by
\begin{equation}
  \label{eq:rhonv}
  \rho_n(v) =  \sum_{|\cmp|=n} \frac{ n  }
{s_\cmp\,s_{\cmp'}}\lp\prod p_i\rp v^l,
\end{equation}
where the sum is taken over all
compositions $\cmp=(p_1,\ldots,p_l)$ of $n$.  Here are the first few
of these polynomials:
\begin{align*}
  \rho_1 & = v \displaybreak[0] \\
  \rho_2 & = v + \frac12 v^2 \displaybreak[0]\\
  \rho_3 & = v + \frac23 v^2 + \frac1{12} v^3 \displaybreak[0]\\
  \rho_4 & = v + \frac34 v^2 + \frac{5}{36} v^3 + \frac{1}{144}v^4
\end{align*}
Routine calculation indicates that the above polynomials can be
factored.  In fact, we have the following general result.
\begin{theorem}
  \label{th:fac1}
  The polynomials $\rho_n(v)$ factorize completely over the
  rationals.  Indeed,
  \begin{equation}
    \label{eq:rhonvfac}
    n!(n-1)!\, \rho_n = v (v+1\cdot 2)(v+2\cdot 3) \ldots
    (v+(n-1)\cdot n).    
  \end{equation}
\end{theorem}
This result can also be appreciated in its converse formulation: the
coefficient of $v^k$ of the Pochammer-symbol-like polynomial in the
right-hand side of \eqref{eq:rhonvfac} is given by the composition sum
\begin{equation}
  \label{eq:csum1}
  \sum_{\subover{|\cmp|=n}{\ell(\cmp)=k}}
  \frac{n!}{s_\cmp}\,\frac{n!}{s_{\cmp'}}\, \prod_{i=1}^k p_i,
\end{equation}
where now the index ranges over only the $k$-part compositions of $n$.
The proof will be given in Section \ref{sect:espot}.  For now let us
stress that the reason underlying this factorization phenomenon is the
exact solvability of the corresponding potential $U(z)$.  We should
also note that this particular $U(z)$ is just a special case of the
well-known Eckart potential.

As regards question 3, for now let us limit ourselves to saying that
at the level of power series the Darboux transformation is given by a
certain invertible, non-linear mapping from a power series
$$U(z)=U_1z+U_2z^2+\ldots$$
to a power series
$$\tU(z)=\tU_1z+\tU_2z^2+\ldots.$$
The mapping in question is described
by a certain composition sum
$$\tU_n = \sum_{|\cmp|=n} w(\cmp) U_\cmp,$$
where the weight $w(\cmp)$
is a rational function of the composition elements $p_1,\ldots,p_l$
that is too complicated to be written down explicitly.  What is
interesting, however, is how easily the Darboux transformation can be
described in terms of spectral residues.
\begin{theorem}
  \label{th:specresdarboux}
  Let $U$ and $\tU$ be formal powers series with zero constant term.
  If $U$ and $\tU$ are related by a Darboux transformation, then the
  corresponding residue sequence of one is the negative of the other.
\end{theorem}
The proof will be given in Section \ref{sect:darboux}.  An interesting
corollary of this theorem (the proof will again be deferred till
Section \ref{sect:darboux} ) is the following inversion formula.  For
each composition $\cmp=(p_1,\ldots,p_{2k+1})$ of odd length set
$$q_\cmp = s_2 s_4 s_6 \ldots s_{2k},$$
where again
$s_j=p_1+\ldots+p_j$.  Next, let $\rho_1,\rho_2,\ldots$ be an infinite
sequence of non-commuting indeterminates, and for each $n=1,2,\ldots$
set
\begin{equation}
  \label{eq:rhotow}
  W_n=\!\!\sum_{\subover{|\cmp|=n}{\ell(\cmp) \text{\rm\ odd}}}
  \frac{1} {(p_1+p_2)(p_2+p_3)\ldots (p_{l-1}+p_l)}\, \rho_\cmp.  
\end{equation}
\begin{corollary}
  \label{cor:winv}
  The inverse mapping is given by
  \begin{equation}
    \label{eq:winv}
  \rho_n = \sum_{\subover{|\cmp|=n}{\ell(\cmp) \text{ odd}}}
  (-1)^{\frac{\ell(\cmp)-1}{2}}\,\frac{1}{q_\cmp\, q_{\cmp'}}
  W_\cmp,    
  \end{equation}
  where $W_\cmp$ is an abbreviation for $W_{p_1}
  W_{p_2}\ldots W_{p_l}$.
\end{corollary}

The organization of the remainder of the paper will mirror the above
introductory discussion.  In Section \ref{sect:specres} we introduce
spectral residues, describe how a potential function determines a
spectral residue sequence, and give the proof of the inversion formula
contained in Theorem \ref{th:rhoinv}.  In Section \ref{sect:espot} we
will treat the spectral residue sequences of the exactly-solvable,
hypergeometric potentials, and derive number of factorization results
similar to Theorem \ref{th:fac1}.  These results are developed in
response to question 2 above, and serve as evidence to the assertion
that exact solvability of a given potential function should manifest
as factorizability of the corresponding spectral residues.  Section
\ref{sect:darboux} we will discuss the Darboux transformation from the
point of view of spectral residue sequences.  The key contents are the
proof of Theorem \ref{th:specresdarboux}, and of the inversion formula
of Corollary \ref{cor:winv}.  In the final section we briefly survey
applications of these identities and inversion formulas to problems in
enumeration and to the theory of integrable systems.

\section{Spectral residues of  Schr\"odinger operators}
\label{sect:specres}
We will begin this section by recalling the details of the classic
method of Frobenius for power series solutions of a second-order
differential equation:
$$z^2 \phi''(z) + P(z) z \phi'(z) + Q(z) \phi(z) = 0.$$
We need to
generalize the classical method somewhat, and therefore allow $P(z)$
and $Q(z)$ to be power series with coefficients in an arbitrary
algebra --- not necessarily commutative --- over the complex numbers.
We want, however, to retain the notion of an indicial equation, and
therefore demand that the constant terms of $P$ and $Q$ be complex
numbers.  It is then easy to show that there exists at least one
formal solution of the form
$$z^\nu ( 1 + \phi_1 z + \phi_2 z^2 + \ldots),$$
where $r=\nu$ is a root of the indicial equation
$$
r^2 + (P_0-1)r +Q_0=0.$$
It is well known that there may not be a
second such solution if the solutions of the indicial equation differ
by an integer.

Since at least one formal solution always exists, no generality will
be lost if we assume that one of the roots is zero.   We will
therefore restrict ourselves to equations 
$$\cT[\phi]=0,$$
where $\cT$ is a differential operator of the form
\begin{equation}
  \label{eq:tdef}
  \cT=z^2 \partial_{zz}+ (1-\nu) z \partial_z + P(z) z\partial_z +
Q(z),  
\end{equation}
and both $P(z)$ and $Q(z)$ are power series with zero constant term.
This way the indicial equation is $r(r-\nu)=0$.

If $\nu$ is not a positive integer, then there is a unique formal
series solution $\phi(z)$ with the initial condition $\phi(0)=1$.  The
coefficients are specified by the following recursive relations:
$$n(\nu-n)\phi_n = Q_n+\sum_{j=1}^{n-1} (jP_{n-j}+Q_{n-j})\phi_j.$$
They can also be described in terms of a composition sum:
\begin{equation}
  \label{eq:phin_comp}
  \phi_n = \sum_{|\cmp|=n} \frac{s_{l-1}P_{p_l}+Q_{p_l}}{n
  (\nu-n)}\times\ldots\times \frac{s_1 P_{p_2}+Q_{p_2}}{s_2(\nu-s_2)}
\times \frac{Q_{p_1}}{s_1 (\nu-s_{1})},  
\end{equation}
where the index ranges over all compositions
$\cmp=(p_1,\ldots,p_l)$ of $n$, and as before
$s_j=p_1+\ldots+p_j$.

When $\nu$ is a positive integer, then in general there is only one
solution; this solution is a series with leading term $z^\nu$.
However, under certain circumstances it is possible for a monic series
solution to exist as well.  To describe when this happens we make the
following definition.
\begin{definition}
  Let $\cT$ be a second-order differential operator as given in
  \eqref{eq:tdef}, but with $\nu$ denoting an indeterminate rather
  than a complex number.  Let $\phi_n$ be the coefficients of the
  corresponding formal monic series solution.  For $n$ a positive
  integer, we define the $n\sth$ spectral residue of $\cT$ to be the
  residue of $\phi_n$ at $\nu=n$.
\end{definition}
\begin{proposition}
  Let $\cT$ be as above, and let $n$ be a positive integer.  A monic
  series solution of 
  $$\cT[\phi]=0\Big|_{\nu=n}$$
  exists if and only if the $n\sth$
  spectral residue of $\cT$ is zero.
\end{proposition}
\begin{proof}
  The $n\sth$ spectral residue vanishes if and only if
  $$Q_n + \sum_{j=1}^{n-1} (jP_{n-j}+Q_{n-j})\phi_j =0
  \Big|_{\nu=n}.$$
  If this is the case, then evidently there exists a
  monic series solution for every possible choice of $\phi_n$.
\end{proof}

Another interesting property of spectral residues is their invariance
with respect to gauge transformations.  In the present context we
define a gauge transformation to be the conjugation of a differential
operator by an invertible multiplication operator.  In particular,
given $\cT$ of the form shown in \eqref{eq:tdef}, and a monic power
series $\mu=1+\mu_1z+\mu_2z^2+\ldots$ we obtain another such
second-order operator by setting
\begin{align*}
\hT &= \mu \circ \cT \circ \mu^{-1}, \\
&=  z^2 \partial_{zz}+ (1-\nu) z \partial_z + \hP(z) z\partial_z +
\hQ(z).
\end{align*}

Gauge transformations arise naturally in the context of linear
differential equations.  Indeed, if $\phi(z)$ is a monic solution of
$\cT[\phi]=0$, then $\hphi=\mu\phi$ is evidently a monic solution of
$\hT[\hphi]=0.$

It is also important to note that the presence of the
$(1-\nu)z\partial_z$ term in $\cT$ will cause the coefficients of
$\hQ(z)$ to depend on the spectral parameter $\nu$.  In the subsequent
discussion we will therefore broaden our class of operators $\cT$ so
as to include the possibility that the coefficients of $P(z)$ and
$Q(z)$ depend polynomially on $\nu$.
\begin{proposition}
  \label{prop:ginv}
  If two second-order differential operators of the form shown in
  \eqref{eq:tdef} are related by a gauge transformation, then all of their
  spectral residues are equal.
\end{proposition}
\begin{proof}
  Let $\cT$, $\hT$ and $\mu$ be as above, and let $\phi(z)$ and
  $\hphi(z)$ be the corresponding monic series solutions.  From
  $\hphi=\mu\phi$ it follows that
  $$\hphi_n = \phi_n + \text{a linear combination of $\phi_1,\ldots
    \phi_{n-1}$}.$$
  From \eqref{eq:phin_comp} it is evident that the
  residues of $\phi_1,\ldots, \phi_{n-1}$ at $\nu=n$ are all zero.
  The desired conclusion follows immediately.
\end{proof}
\noindent

A particularly important type of second-order differential operator is
the class of Schr\"odinger operators (the term Hamiltonian operator
will also be used.)  In the present setting this will be taken to mean
an operator of the form
\begin{equation}
  \label{eq:sopdef}
  \cH = z^2\partial_{zz} + (1-\nu) z\partial_z + U(z), 
\end{equation}
where the potential function $U(z)=U_1z+U_2 z^2+\ldots$ is a formal
series with vanishing constant term, and where again we allow the
coefficients $U_1, U_2, \ldots$ to depend polynomially on $\nu$.

It is more customary to define a Schr\"odinger operator as a formally
self-adjoint operator of the form
\begin{equation}
  \label{eq:fullsa}
  \hH=z^2\partial_z + z\partial_z+U(z)-\lambda,  
\end{equation}
where $\lambda$ is
the spectral parameter, and $z$ is related to the physical distance
variable $x$ by $z=e^x$.  However, the formal eigenfunctions of $\hH$
are of the form $z^{\pm\sqrt\lambda}$ times a monic series.  To get at
the monic series directly we conjugate $\hH$ by $z^{\nu/2}$, with
$\lambda=\nu^2/4$, to obtain an operator in the ``nearly''
self-adjoint form shown in \eqref{eq:sopdef}.

It is important to note that the class of Schr\"odinger operators
provides a canonical form for second-order differential operators with
respect to gauge-transformations.
\begin{proposition}
  \label{prop:gsop}
  Let $\cT$ be a second-order operator of the form shown
  in \eqref{eq:tdef}.
  There exists a monic series
  $\mu(z)$ and a potential function
  $U(z)$ with zero constant term such that
  $$z^2\partial_{zz} + (1-\nu) z\partial_z + U(z) = \mu \circ \cT
  \circ \mu^{-1}.$$ 
\end{proposition}
\begin{proof}
  In the non-commutative setting it suffices to take
  \begin{align*}
    \mu_n &= \frac{1}{2n}\lp P_n + \sum_{j=1}^{n-1} \mu_{n-j} P_j \rp,\\
    U &= \lp \mu Q  - z^2\mu'' -(1-\nu)\mu' \rp \mu^{-1},
  \end{align*}
  where $P(z)$, $Q(z)$ are the coefficient series of the operator $\cT$.
  In the commutative setting one can use simpler formulas.  Indeed, it
  suffices to introduce an auxiliary function
  $$\sigma(z) = \int \frac{P(z)}{2z},\quad \sigma(0)=0,$$
  and then set
  \begin{align}
    \nonumber
    \mu &= e^{\sigma},\\
    \label{eq:usigmarel}
    U &= Q - z^2 \sigma'' - (1-\nu)z\sigma' - z^2(\sigma')^2.
  \end{align}
\end{proof}

Gauge invariance of the spectral residues will be of crucial
importance when we consider hypergeometric potentials in the next
section.  For the remainder of the present section, we will focus our
attention on the spectral residues of a Schr\"odinger operator.

\begin{proposition}
  \label{prop:specres}
  The $n\sth$ spectral residue, call it $\rho_n$, of a Schr\"odinger
  operator with potential $U(z)=U_1z+U_2z^2+\ldots$ is given by
  $$
  \rho_n = \sum_{|\cmp|=n} \frac{n}{s_\cmp\,s_{\cmp'}} \,U_\cmp.
  $$
\end{proposition}
\begin{proof}
  Taking $P(z)=0$ and $Q(z)=U(z)$, the desired conclusion follows
  immediately from \eqref{eq:phin_comp} above.
\end{proof}

From formula \eqref{eq:rhodef} we see that $\rho_n = n^{-1} U_n +
$ lower order terms.  It is therefore evident that the sequence of
spectral residues $\rho_1, \rho_2, \ldots$ completely determines the
potential.  Surprisingly, the formula for the inverse mapping can be
given explicitly.  This is formula \eqref{eq:rhoinv} of Theorem
\ref{th:rhoinv}.  We conclude the present section by giving a proof of
this theorem.

Let $\cH$ be a Schr\"odinger operator of the form shown in
\eqref{eq:sopdef}, and let $\phi(z)=1+\phi_1 z + \phi_2 z^2 + \ldots$
be the formal monic series solution of 
$$\cH[\phi]=0.$$
Specializing 
\eqref{eq:phin_comp} we see that 
\begin{equation}
  \label{eq:sopsol}
  \phi_n = \sum_{|\cmp|=n} \frac{U_{p_l}}{n
  (\nu-n)}\times\ldots\times \frac{U_{p_2}}{s_2(\nu-s_2)}
\times \frac{U_{p_1}}{s_1 (\nu-s_{1})}.  
\end{equation}
For all pairs of positive integers $k$ and $n$ with $k\leq n$, we let
$\rho_{n,k}$ denote the residue of $\phi_n$ at $\nu=k$.  Thus,
$\rho_{n,n}$ is just another way to refer to the $n\sth$ spectral
residue $\rho_n$.

We are now able to write
\begin{equation}
  \label{eq:phin_rhosum} 
  \phi_n = \frac{\rho_{n,n}}{\nu-n} + \frac{\rho_{n,n-1}}{\nu-n+1} +
  \ldots + \frac{\rho_{n,1}}{\nu-1}. 
\end{equation}

Next, we evaluate $\rho_{n,k}$ in terms of the coefficients of
$U(z)$.  From \eqref{eq:sopsol} we obtain
$$\rho_{n,k} = \!\!\!\sum_{\subover{|\cmp|=k}{|\cmpq|=n-k}}
\frac{U_{q_m} \ldots U_{q_1}}{(k+t_m)(-t_m)\ldots (k+t_1)(-t_1)}
\times \frac{U_{p_l} \ldots U_{p_1}}{k \ldots (k-s_2)s_2(k-s_1)s_1},$$
where the sum is indexed by all compositions $\cmp=(p_1,\ldots,p_l)$
of $k$ and all compositions $\cmpq=(q_1,\ldots,q_m)$ of $n-k$, and
where $s_j=p_1+\ldots+p_j$ and $t_j=q_1+\ldots+q_j$.  It is evident by
inspection that the above sum can be factored, and indeed that
$$\rho_{n,k} = \phi_{n-k}\Big|_{\nu=-k}\!\!\! \times \rho_k.$$
Using \eqref{eq:phin_rhosum} once again we obtain
\begin{equation}
  \label{eq:rhonk_rec}
  \rho_{n,k} = - \lp \frac{\rho_{n-k}}{k+n} +
  \frac{\rho_{n-k,n-k-1}}{k+n-1} + \ldots + \frac{\rho_{n-k,1}}{k+1}\rp
  \rho_k.  
\end{equation}

Now $\phi$ is a solution of $\cH[\phi]=0$ if and only if 
$$n(\nu-n) \phi_n = U_n + \sum_{j=1}^{n-1} U_{n-j} \phi_j.$$
Substituting \eqref{eq:phin_rhosum} into the left hand side, and
evaluating both sides at $\nu=\infty$ we obtain
$$
U_n = n\lp\rho_n + \rho_{n,n-1} + \ldots + \rho_{n,1}\rp.
$$
This equation together with \eqref{eq:rhonk_rec} gives a recursive
specification of $U_n$ in terms of $\rho_1,\ldots,\rho_n$.  Rewriting
this as a composition sum we obtain the desired formula
\eqref{eq:rhoinv}. \qed

\section{Spectral residues of exactly solvable potentials}
\label{sect:espot}
We have already mentioned in the introduction that the spectral
residue sequence is sensitive to the exact solvability of the
underlying potential.  In the present section we will discuss this
phenomenon in some detail, and use it to derive a number of
factorization identities similar to the identity described in Theorem
\ref{th:fac1} above.

The key to the whole matter is the invariance of spectral residues
with respect to gauge transformations --- see Proposition
\ref{prop:ginv} above.  Hypergeometric functions also play an
important role.  Indeed the exact solvability of the potentials that
we will be considering here is based on the fact that the
corresponding Schr\"odigner operators are gauge equivalent to the
hypergeometric operator.  Certain values of the hypergeometric
parameters cause the hypergeometric series to truncate, which means
that for those values of the parameters almost all the spectral
residues vanish.  Thus, when we consider spectral residues as
polynomials in the hypergeometric parameters we obtain factorizations
of said polynomials.

Let us also remark that in the present section we can safely restrict
ourselves to power series with complex coefficients.  We will consider
non-commutative coefficients again in section \ref{sect:darboux}.

Three families of exactly-solvable potentials will be treated here:
the Eckart \cite{Eckart}, the P\"oschl-Teller \cite{PoschlTeller}, and
the Morse \cite{Morse} potentials.  The first two of these have
eigenfunctions related to Gaussian hypergeometric functions; in the
latter case one needs to use confluent hypergeometric functions.

We will discuss each of these families in turn. In each case we will
first show how the Schr\"odinger's equation is related to the
hypergeometric or the confluent hypergeometric equation.  Next we will
consider the corresponding spectral residues and see that in each case
there is complete factorization in terms of the potential parameters.
Finally, we will list some interesting factorization identities that
arise for each of the families for certain specialized values of the
parameters.

\subsection{Eckart Potentials.}
The Eckart potentials are functions of the form
$$U=\frac{u\, \E^x}{1-\E^x}+\frac{v\,\E^x}{(1-\E^x)^2},$$
where $x$ is
the physical distance variable, and $u$ and $v$ are parameters.
Rewriting the Hamiltonian, $\cH$, in terms of $z=\E^x$ we get
\begin{equation}
  \label{eq:eckartham}
  \cH=z^2\partial_{zz}+(1-\nu)z\partial_z+\frac{uz}{1-z}+\frac{vz}{(1-z)^2},  
\end{equation}
where the spectral parameter $\nu$ has been encoded into the
first-order part of the operator --- see the preceding Section for a
discussion of this matter.

Let us now look for the connection between Eckart potentials and
hypergeometric functions.  The hypergeometric series,
$$F(\alpha,\beta,\gamma;z)=F_0+F_1z+F_2z^2+\ldots,$$
can be defined as
the solution of the hypergeometric differential equation:
\begin{equation}
  \label{eq:Fsol}
  \cG(F)=0,\quad F(0)=1,
\end{equation}
where $\cG$ is the familiar second order differential operator
\begin{equation}
  \label{eq:hgop}
  \cG = z(1-z) \partial_{zz} + \lp\gamma-(\alpha+\beta+1)z\rp
  \partial_z-\alpha\beta.  
\end{equation}

In order to work with spectral residues we need to somehow convert the
hypergeometric operator into an operator of the form $z^2\partial_{zz}
+ $ lower order terms.  An obvious way to do this is to left-multiply
by $z/(1-z)$.  To wit, set
\begin{eqnarray*}
\cG_1&=&\frac{z}{1-z}\,\cG\\
&=& z^2\partial_{zz}+(1-\nu)z\partial_z -
\frac{z}{1-z}\lb (\alpha+\beta+\nu) z\partial_z + \alpha\beta\rb
\end{eqnarray*}
where $\nu=1-\gamma$ is the required spectral parameter.  The upshot
of all this is that one can equally well specify the hypergeometric
series $F$ as the solution of
$$\cG_1(F)=0,\quad F(0)=1.$$

In Proposition \ref{prop:gsop} of the preceding section we noted that
every second-order differential operator is gauge-equivalent to a
Schr\"odinger operator.  Following the method described there, we
conjugate $\cG_1$ by a multiplication operator $\mu=e^\sigma$ where
\begin{align*}
  \sigma(z) &= -\frac12(\alpha+\beta+\nu)\int \frac{1}{1-z}  \\
  &= \frac 12 (\alpha+\beta+\nu)\, \log(1-z) ;\\
  \mu(z) &= (1-z)^{(\alpha+\beta+\nu)/2}.
\end{align*}
The end-result of the gauge-transformation is the Eckart Hamiltonian shown in
\eqref{eq:eckartham}.  Using \eqref{eq:usigmarel}, an elementary calculation
will show that the potential parameters are related to the hypergeometric
parameters by
\begin{align}
  \label{eq:uvabrel}
  u&=\frac14\,(\alpha-\beta-\nu)(\alpha-\beta+\nu), \\
  \nonumber
  v&=\frac14\,(\alpha+\beta+\nu)(2-\alpha-\beta-\nu).
\end{align}

Turning next to spectral residues, we expand the Eckart potential in a series
with respect to $z$:
$$U=\sum_k (u+kv) z^k.$$
In Proposition \ref{prop:specres} we proved
that the spectral residues of a Schr\"odinger equation are given by
formula \eqref{eq:rhodef}.  From this formula and from the above
expansion we obtain the expression for the spectral residues of the
Eckart potentials:
\begin{equation}
  \label{eq:pdef}
\rho_n(u,v) = n \sum_{|\cmp|=n} \frac{
\prod_{i} (u+vp_i)}
{s_\cmp\,s_{\cmp'}},
\end{equation}
where as usual the index ranges over all compositions
$\cmp=(p_1,\ldots,p_l)$ of $n$.  We have arrived at the following
result.
\begin{proposition}
  \label{prop:eckartfac}
  In the even case, $n=2m$, one has
  \begin{equation}
    \label{eq:ecfaceven}
    \rho_n(u,v)= K \prod_{i=0}^{m-1} \left[ (u+v+a_{n,i})(u+v+a_{n,i+1}) +
      \bisq v\right],
  \end{equation}
  where $K=\frac1{n!(n-1)!}$, $a_{n,i}=i(n-i)$ and $b_{n,i} = n-1-2i$.  In the
  odd case, $n=2m+1$, one has
  \begin{equation}
    \label{eq:ecfacodd}
    \rho_n(u,v)= K (u+v+a_{n,m}) \prod_{i=0}^{m-1} \left[
      (u+v+a_{n,i})(u+v+a_{n,i+1}) + \bisq v\right].
  \end{equation}
\end{proposition}
\begin{proof}
  We saw above that the eigenfunctions of the Eckart Hamiltonian are
  the product of a power of $(1-z)$ and of a hypergeometric series.
  Hence by the gauge-invariance of spectral residues as asserted in
  Proposition \ref{prop:ginv} above, the spectral residues of the
  Eckart Hamiltonian can just as well be calculated by taking residues
  of the coefficients of the hypergeometric series.
  
  These are of course the well-known:
  \begin{equation}
    \label{eq:hgcoeffs}
    F_n = \frac{[\alpha]^n [\beta]^n}{n!\, [\gamma]^n},
  \end{equation}
  where $[x]^n$ denotes the rising factorial $x(x+1)\ldots(x+n-1)$.
  Calculating the residue of $F_n$ at $\nu=n$ (recall that
  $\gamma=1-\nu$), and using the gauge-invariance of $\rho_n$ we
  obtain 
  \begin{equation}
    \label{eq:abres}
    \rho_n(u,v)\Big|_{\nu=n}= (-1)^n
    \frac{[\alpha]^n [\beta]^n}{n!(n-1)!}.
  \end{equation}
  From \eqref{eq:uvabrel} it follows that
 \begin{multline}
    \label{eq:uvfactors}
    (\alpha+i)(\beta+i)
    (\alpha+n-1-i)(\beta+n-1-i)= \\
    (u+v+a_{n,i}) (u+v+a_{n,i+1}) + \bisq v \; \Big|_{\nu=n} ,
  \end{multline}
  Another calculation will show that if $n$ is odd, say $n=2m+1$, then
  $$
  -(\alpha+m)(\beta+m)=u+v+a_{n,m}\; \Big|_{\nu=n} .$$
  This observation explains why
  the $(-1)^n$ factor in \eqref{eq:abres} can be discarded.  The proposition
  now follows immediately.
\end{proof}


To conclude the discussion of the Eckart potentials let us mention
some identities that come about as a consequence of Proposition
\ref{prop:eckartfac}.  The first of these was mentioned in Theorem
\ref{th:fac1} of the introduction.  For the proof, we simply set $u=0$
in \eqref{eq:ecfaceven} and \eqref{eq:ecfacodd} above, and note that
$$(v+a_{n,i})(v+a_{n,i+1})+\bisq v = \lp v+i(i+1) \rp \lp v
+(n-i-1)(n-i) \rp.$$
A similar factorization identity can also be
obtained by setting $u=-v$ in \eqref{eq:ecfacodd} and
\eqref{eq:ecfaceven}.  In order to describe this identity, for
positive integers $k\leq n$ let us set
\begin{equation}
  \label{eq:csum2} 
c_{n,k} = \sum_{\subover{|\cmp|=n}{\ell(\cmp)=k}}
\frac{n!}{s_\cmp}\frac{n!}{s_{\cmp'}} \prod_i (p_i-1)
\end{equation}
\begin{corollary}
  In the even case, $n=2m$, the above numbers $c_{n,k}$ are given by
  the coefficient of $x^k$ in
  $$
  \prod_{i=0}^{m-1} \Big( (n-1-2i)^2x +i(i+1)(n-1-i)(n-i)\Big).
  $$
  An analogous statement holds in the odd case, $n=2m+1$, but with
  an additional factor of $m(m+1)$ appended to the above expression.
\end{corollary}


\subsection{P\"oschll-Teller Potentials.}
Turning next to the P\"oschl-Teller potentials, these are functions of
the form
$$ \frac{u}{\cosh^2(x/2)} + \frac{v}{\sinh^2(x/2)},$$
where again $x$ is the physical distance variable, and $u$, $v$ are
parameters.  
Rewriting the Hamiltonian, $\cH$, in terms of $z=\E^x$ we get
\begin{equation}
  \label{eq:ptham}
  \cH=z^2\partial_{zz}+(1-\nu)z\partial_z+\frac{uz}{(1+z)^2}+
  \frac{vz}{(1-z)^2}. 
\end{equation}

Let us next consider how $\cH$ can be related to the hypergeometric
operator $\cG$.  To begin, let us rewrite the latter using $\zeta$ as
the independent variable:
$$
\cG = \zeta(1-\zeta) \partial_{\zeta\zeta} +
\lp\gamma-(\alpha+\beta+1)\zeta\rp \partial_\zeta-\alpha\beta.
$$
We note that a change of variables $\cosh x=2\zeta-1$ will
transform the leading term of $\cG$ into $-\partial_{xx}$.  Since we
would like the leading term to be $-z^2\partial_{zz}$ (the minus sign
doesn't really matter), we employ $z=e^x$ as the independent variable.
To make the change of variables we note that
\begin{equation}
  \label{eq:zzetarel}
  \zeta=\frac{(z+1)^2}{4z},\quad
  \partial_\zeta=\frac{4z^2}{z^2-1}\partial_z.  
\end{equation}
In this way we obtain
\begin{multline*}
  -\cG = z^2\partial_{zz} +(1-\alpha-\beta)z\partial_z+ \\
  \lp(2\gamma-1)\frac{z}{z+1}+(2\alpha+2\beta+1-2\gamma)\frac{z}{z-1}\rp
  z\partial_z+ \alpha\beta
\end{multline*}
The roots of the indicial equation are $\alpha$ and $\beta$, and
therefore, in order to obtain solutions that are monic power series we
make a gauge-transformation:
\begin{align*}
  \cG_1&=-z^{-\alpha}\circ \cG \circ z^\alpha \\
  &= z^2\partial_{zz}+(1+\alpha-\beta)z\partial_z+ \\
  &\qquad +\lp(2\gamma-1)\frac{z}{z+1}+
  (2\alpha+2\beta+1-2\gamma)\frac{z}{z-1}\rp ( z\partial_z + \alpha)
\end{align*}
Employing the abbreviations
\begin{align}
  \label{eq:newhgparms}
  \lambda&=\gamma-1/2,\\
  \mu&=\alpha+\beta+1/2-\gamma,\nonumber\\
  \nu&=\beta-\alpha,\nonumber\\
  w(z)&=\frac{\lambda z}{z+1} +\frac{\mu z}{z-1}\nonumber
\end{align}
the above operator can be more compactly rewritten as
$$
\cG_1=z^2\partial_{zz}+(1-\nu)z\partial_z+ 2w(z) z \partial_z +
(\lambda+\mu-\nu)w(z)
$$
From there, the following gauge transformation will give the
P\"oschl-Teller Hamiltonian:
$$\cH = \lp(1+z)^\lambda(1-z)^\mu\rp\circ \cG_1\circ \lp
(1+z)^{-\lambda}(1-z)^{-\mu}\rp,$$
provided the potential parameters
are related to the hypergeometric parameters by
$$ u=\lambda^2-\lambda,\quad v=\mu-\mu^2.$$

Turning next to the eigenfunctions of $\cH$ we encounter a difficulty:
we do not get a power series if we substitute $\zeta=(z+1)^2/(4z)$
into the usual hypergeometric series $F(\alpha,\beta,\gamma;\zeta)$.
Note though that $4z/(1+z)^2$ is analytic in $z$, and so we can
overcome this difficulty by employing a solution of the hypergeometric
equation that is analytic, or nearly so, in $\zeta^{-1}$.  Just such a
solution can be found on Kummer's list of 24 solutions to the
hypergeometric equation --- see Section 2.9 of \cite{Bateman}.  The
solution to $\cG(\varphi)=0$ we require is
$$
\varphi(\zeta)=\zeta^{-\alpha}
F(\alpha,\alpha+1-\gamma,\alpha+1-\beta;\zeta^{-1}).
$$
Let us summarize this observation in the following manner.
\begin{lemma}
  \label{lemma:phisol}
  The unique series solution $\phi=\phi(z)$ to
  $$\cG_1(\phi)=0,\quad \phi(0)=1,$$
  is given by
  $$
  \phi(z) = (1+z)^{\nu-\lambda-\mu}
  F\lp\frac{\mu+\lambda-\nu}{2}\,,\frac{\mu-\lambda+1-\nu}{2}\,,1-\nu\,;
  \frac{4z}{(1+z)^2}\rp.
  $$
\end{lemma}

We are now in a position to factorize the spectral residues of the
P\"oschl-Teller potential.  Expanding the P\"oschl-Teller potential
in a power series we obtain
$$U(z)=\sum_{k\text{ even}} ku_0\,z^k+\sum_{k\text{ odd}} ku_1\,z^k,$$
where $u_0=v-u$ and $u_1=v+u$.  Hence, by Proposition
\ref{prop:specres} the $n\sth$ spectral residue, is given by
\begin{equation}
  \label{eq:ptrhodef}
  \rho_n = n \sum_{|\cmp|=n} \frac{\prod_i p_i}{s_\cmp s_{\cmp'}} \,
  u_1^{\nodd(\cmp)} u_0^{\neven(\cmp)},
\end{equation}
where $\nodd(\cmp)$ is the
number of odd elements of a composition $\cmp=(p_1,\ldots,p_l)$, and
$\neven(\cmp)$ is the number of even elements.
\begin{proposition}
  \label{prop:ptfac}
  In the even case, $n=2m$, one has
  \begin{equation}
    \rho_n= K\!\!\!\prod_{\subover{0<k<n}{k\text{\rm\
    odd}}}\!\!\!\! \Big( u_1^2+2k^2u_0+k^4-k^2\Big), 
  \end{equation}
  where $K=\frac{1}{n!(n-1)!}$.  In the odd case, $n=2m+1$, one has
  \begin{equation}
    \rho_n= Ku_1\!\!\!
    \prod_{\subover{0<k<n}{k\text{\rm\ even}}}\!\!\!\!\Big(
    u_1^2+2k^2u_0+k^4-k^2\Big), 
  \end{equation}
\end{proposition}
\begin{proof}
  Let $\phi(z)=1+\phi_1 z+\phi_2 z^2+\ldots$ be the unique series
  solution to
  $$\cG_1(\phi)=0,\quad \phi(0)=1.$$
  By Lemma \ref{lemma:phisol}, if
  $(\mu+\lambda-\nu)/2=-j$, $j\in\natnums$, then $\phi(z)$ is
  $(1+z)^{\nu-\lambda-\mu}$ times a $j\sth$ degree polynomial of
  $z/(z+1)^2$ whose coefficients are polynomials of $\lambda, \mu,
  \nu$ over $(1-\nu)(2-\nu)\ldots(j-\nu)$.  Hence for all $n>j$ the
  residue of $\phi_n$ at $\nu=n$ must have $\mu+\lambda+2j-n$ as a
  factor. By the gauge invariance of spectral residues, the same must
  be true for $\rho_n$.  The same reasoning also applies to show that
  $\lambda-\mu-1-2j+n$ is a factor of $\rho_n$ for all
  $j=0,1,\ldots,n-1$.
  
  We have now shown that $\rho_n$, as a function of $\lambda$ and
  $\mu$, is some factor $K$ times
  $$\prod_{j=0}^{n-1} (\mu+\lambda+2j-n)(\lambda-\mu-1-2j+n).$$
  Now
  \begin{align*}
    u_0 &= v-u=\mu+\lambda-\mu^2-\lambda^2,\\
    u_1 &=
    v+u=\mu-\lambda-\mu^2+\lambda^2=(\lambda-\mu)(\mu+\lambda+1),
  \end{align*}
  and hence for all $k$
  \begin{multline*}
    u_1^2+2k^2u_0 +k^4-k^2 = \\
    (\lambda+\mu+k-1)(\lambda+\mu-k-1)(\lambda-\mu+k)(\lambda-\mu-k)
  \end{multline*}
  Using \eqref{eq:ptrhodef}, a degree count, and an examination of the
  coefficient of $u_1^n$ we see that $K=\frac{1}{n!(n-1)!}$. The
  proposition now follows immediately.
\end{proof}

Here are some interesting identities arising from the preceding
factorization.  
\begin{corollary}
  The coefficient of $x^k$ in
  $$(x+2^4)(x+4^4)(x+6^4)\ldots(x+(2m)^4)$$
  is given by the following composition sum:
  \begin{equation}
    \label{eq:corrid}
    \sum_{\subover{|\cmp|=2m+1}{\nodd(\cmp)=2k+1}}\!\!\!
    \frac{n!}{s_\cmp}\frac{n!}{s_{\cmp'}}
    \frac{\prod_i  p_i}{2^{\neven(\cmp)}}
  \end{equation}
\end{corollary}
\begin{proof}
  Upon setting $x=u_1^2$, $u_0=1/2$, the above identity follows from
  the odd, $n=2m+1$, case of Proposition \ref{prop:ptfac}.
\end{proof}
If we use the same summand as above, but restrict ourselves to
compositions without even numbers we obtain the coefficients of a
closely related product.
\begin{corollary}
  The coefficient of $x^k$ in
  $$(x+2^4-2^2)(x+4^4-4^2)(x+6^4-6^2)\ldots(x+(2m)^4-(2m)^2)$$
  is given by the following compositions sum:
  $$
  \sum_{
    \stackrel{ |\cmp|=2m+1}
             { \stackrel{\nodd(\cmp)=2k+1}
                        {\scriptscriptstyle\neven(\cmp)=0}
             }}
  \frac{n!}{s_\cmp}\frac{n!}{s_{\cmp'}} \prod_i p_i
  $$
\end{corollary}
\begin{proof}
  Upon setting $x=u_1^2$, $u_0=0$, the above identity follows from the
  odd, $n=2m+1$, case of Proposition \ref{prop:ptfac}.
\end{proof}
Identities featuring $k^4$ and $k^4-k^2$, with $k$ odd, can be
obtained in an analogous manner.

\subsection{Morse Potentials.}
Finally, let us consider Morse potentials;  these are functions of the
form 
$$ue^x + ve^{2x},$$
where $x$ is the physical distance variable, and
$u$, $v$ are parameters.  Rewriting the Hamiltonian, $\cH$, in terms
of $z=\E^x$ we get
\begin{equation}
  \label{eq:morseham}
  \cH=z^2\partial_{zz}+(1-\nu)z\partial_z+uz+vz^2,  
\end{equation}

The eigenfunctions of the above operator can be related to confluent
hypergeometric functions by a gauge transformation.  The confluent
hypergeometric series can be defined as the solution of the confluent
hypergeometric differential equation:
$$  \cJ(\phi)=0,\quad \phi(0)=1,$$
where $\cJ$ is the second order differential operator
$$
\cJ = z\partial_{zz} + (\gamma-z)\partial_z + \alpha $$
We need to
introduce an extra scaling parameter, $\omega$, stemming from the
change of variables $z\mapsto -2\omega z$.  Now, the relevant operator
is given by
$$\cJ = z\partial_{zz} + (\gamma+2\omega z)\partial_z + 2\omega \alpha,$$
and the solution, 
$$\phi(\alpha,\gamma,\omega;z)=\phi_0+\phi_1z+\phi_2z^2+\ldots,$$
of $\cJ[\phi]=0$ depends on 3 parameters.

The equation $\cJ[\phi]=0$ is equivalent to a certain
relation between $\phi_n$ and $\phi_{n+1}$, and yields the following
modification of the familiar formula for the coefficients of $\phi$:
$$\phi_n = \frac{(-2)^n \omega^n [\alpha]^n}{n!\,[\gamma]^n}.$$

Proceeding as we did in the case of the Eckart potentials we modify
$\cJ$ so as  to obtain an operator with leading term
$z^2\partial_{zz}$:
$$
  \cJ_1=z\cJ = z^2\partial_{zz}+(1-\nu)z\partial_z + 2\omega z^2\partial_z +
  2\omega \alpha z,  
$$
where again we employ $\nu=1-\gamma$.  Next we convert $\cJ_1$ to
(nearly) self-adjoint form by means of a gauge transformation with
gauge factor $\mu=e^\sigma$, where
$$
  \sigma = \int^z \omega = \omega z.
$$
Hence, 
$\mu  = e^{\omega z}.$
The result is the Hamiltonian operator shown in \eqref{eq:morseham},
where the potential parameters are related to the hypergeometric
parameters by
\begin{align}
  \label{eq:muvabrel}
  u&=\omega(\nu+2\alpha-1),\\
  \nonumber
  v&=-\omega^2 .
\end{align}
To obtain the corresponding spectral residues we again employ
\eqref{eq:rhodef}:
$$
\rho_n(u,v) = n\sum_{\stackrel{|\cmp|=n} {\scriptscriptstyle
    p_i\in\{1,2\}}} \frac{
  u^{\nodd(\cmp)}\,v^{\neven(\cmp)}}{s_\cmp\,s_{\cmp'}},
$$
where the index now ranges over compositions
$\cmp=(p_1,\ldots,p_l)$ of $n$ consisting of $1$ and $2$ only, and
where it is necessary to recall that $\nodd(\cmp)$ denotes the number
of odd elements of a composition $\cmp=(p_1,\ldots,p_l)$, and
$\neven(\cmp)$ the number of even elements.

\begin{proposition}
  \label{prop:morsefac}
  In the even case, $n=2m$, one has
  \begin{equation}
    \label{eq:morsefaceven}
    \rho_n(u,v)= K\!\!\!\prod_{\subover{0<k<n}{k\text{\rm\
    odd}}}\!\!\! (u^2+k^2v ),
  \end{equation}
  where $K=\frac{1}{n!(n-1)!}$.  In the odd case, $n=2m+1$, one has
  \begin{equation}
    \label{eq:morsefacodd}
    \rho_n(u,v)= Ku\!\!\!\prod_{\subover{0<k<n}{k\text{\rm\
    even}}}\!\!\! (u^2+k^2v ),
  \end{equation}
\end{proposition}
\begin{proof}
  We proceed in the same way as in the proof of Proposition
  \eqref{prop:eckartfac}.  By the gauge-invariance of spectral
  residues, $\rho_n$ is equal to the $n\sth$ spectral residue of the
  confluent hypergeometric operator.  Taking the residue of $\phi_n$
  at $\nu=n$ yields
  $$\frac{2^n\omega^n[\alpha]^n}{(n!)^2}.$$
  Next, using
  \eqref{eq:muvabrel} we see that in the first case, where $n=2m,
  k=2i+1$, we have
  $$u^2+k^2 v\;\Big|_{\nu=n} \!\!\!=
  4\,\omega^2(\alpha+m+i)(\alpha+m-i-1).$$
  In the second case, where $n=2m+1$, $k=2i$ we have
  \begin{align*}
  u^2+k^2 v\;\Big|_{\nu=n} \!\!\!&=
  4\,\omega^2(\alpha+m+i)(\alpha+m-i),\\
  u\;\Big|_{\nu=n} \!\!\!&= 2\,\omega(\alpha+m).
  \end{align*}
  The proposition now follows immediately.  
\end{proof}
Setting $v=1$ in the above Proposition immediately yields the
following identities:
\begin{corollary}
  The coefficient of $x^k$ in 
  $$(x+2^2)(x+4^2)(x+6^2)\ldots(x+(2m)^2)$$
  is given by the following composition sum:
  $$
  \sum_{
    \stackrel{ |\cmp|=2m+1}
             { \stackrel{\nodd(\cmp)=2k+1}
                        {\scriptscriptstyle p_i\in\{1,2\}}
             }}
  \frac{n!}{s_\cmp}\frac{n!}{s_{\cmp'}}
  $$
  The coefficient of $x^k$ in 
  $$(x+1^2)(x+3^2)(x+5^2)\ldots(x+(2m-1)^2)$$
  is given by the following composition sum:
  $$
  \sum_{
    \stackrel{ |\cmp|=2m}
             { \stackrel{\nodd(\cmp)=2k}
                        {\scriptscriptstyle p_i\in\{1,2\}}
             }}
  \frac{n!}{s_\cmp}\frac{n!}{s_{\cmp'}}
  $$
\end{corollary}


\section{The Darboux transformation}
\label{sect:darboux}
We begin this section by describing the Darboux transformation for
linear, second-order differential equations.  Let us consider a
self-adjoint, second-order operator
$$\cH=\partial_{xx} + U(x),$$
and ask whether it can be factored into a product of first order
operators.   It is not hard to see that the answer is ``yes'', and
that the factorization must be of the form
$$\cH = (\partial_x - W) (\partial_x + W),$$
where the function
$W(x)$ --- we will refer to it as a prepotential --- is related to the
potential by $U = W_x - W^2$.  Reversing the order of the above
factorization we obtain another self-adjoint operator:
$$\tH = (\partial_x + W) (\partial_x - W) = \partial_{xx} + \tU,$$
where now $\tU = -W_x - W^2$.  Knowing $U$ one has to solve a Ricatti
equation in order to obtain $W$.  Alternatively, one can describe the
prepotential as $W=-\partial_x(\log \psi)$, where $\psi(x)$ is a
solution of $\cH(\psi)=0$.

The process of going from $U(x)$ to $\tU(x)$ is called the Darboux
transformation. It is evident, that at the level of the prepotential,
the Darboux transformation corresponds to a negation, and that
therefore the Darboux transform of $\tU$ takes us back to $U$.

There is a close relationship between the (formal) eigenfunctions of $\cH$
and $\tH$.  Indeed, one can write
\begin{align}
  \label{eq:itwine}
(\partial_x+W)(\partial_x - W) (\partial_x + W)
&= (\partial_x+W) \cH\\
\nonumber  &= \tH (\partial_x + W), 
\end{align}
From here, suppose that $\psi$ is a formal eigenfunction of $\cH$,
i.e. that
$\cH(\psi) = \lambda \psi$.
Setting $\tpsi=\psi_x + W\psi$ we employ \eqref{eq:itwine} to conclude
that $\tH(\tpsi) = \lambda \tpsi$.

In order to apply the Darboux transformation in a truly analytic
setting, one has to take into account factors like square-summability
of the eigenfunctions, and smoothness of the potential.  Here, we are
more interested in formal properties of the Darboux transformation.
Moreover, our approach is based on power series in $z=e^x$, rather
than on functions of $x$.

Let us therefore adapt the notion of a Darboux transformation to the
present setting.  We consider a second order operator
$$\cH=z^2\partial_{zz} + (1-\nu)z\partial_z + U,$$
where $U(z)$ is a
formal power series with vanishing constant term.  We do not assume
that the coefficients $U_n$ commute.  Owing to the fact that $U(z)$
has zero constant term, $\cH$ cannot be factored into a product of
first order operators.  In order to obtain a factorization we must add
a constant.  To wit,
$$\cH+\nu^2/4=(z\partial_z -\nu/2 - W)(z\partial_z -\nu/2 + W),$$
where $W(z)=W_1z +W_2 z^2 + \ldots$ is another formal power series
related to the potential by 
$$U = zW_z - W^2.$$
At the level of coefficients this is equivalent to an infinite number 
of graded relations:
$$
U_n = n W_n - \sum_{i=1}^{n-1} W_i W_{n-i}.$$
Note that $W_n =
n^{-1}\, U_n + $ lower order terms, and hence the above relations can be
recursively inverted, so that if we know the coefficients of $U$, we
can calculate the coefficients of $W$.

Reversing the above factorization we arrive at the partner operator 
$$\tH = z^2\partial_{zz} + (1-\nu)z\partial_z + \tU,$$
were $\tU = -z W_z - W^2.$
Here are the first few relations that describe the
formal Darboux transformation:
$$
\tU_1 = - U_1,\quad
\tU_2 = -2\,U_1^{\,2}-U_2,\quad
\tU_3 = -2\,U_1^{\,3}-2\,U_1U_2-U_3
$$
It does not seem possible to obtain a closed sum formula for the
above relations.

Next, let $\phi(z) = 1 + \phi_1 z + \phi_2 z^2 + \ldots$ be the formal
monic solution of 
$$\cH(\phi)=0.$$
We saw in Section
\ref{sect:specres} that the coefficients of $\phi$ will be certain
rational functions in $\nu$.  We set
$$\tphi = z\phi_z + (-\nu/2 +W)\phi,$$ 
and note that since
$$(\tH+\nu^2/4)(z\partial_z-\nu/2+W) =
(z\partial_z-\nu/2+W)(\cH+\nu^2/4),$$
we can conclude that
$$\tH(\tphi) = 0.$$

We are now ready to prove Theorem \ref{th:specresdarboux}.  Note
that 
$$\psi=-2\, \nu^{-1} \tphi$$
is the unique monic solution of
$$\tH(\psi)=0.$$
Hence the $n\sth$ spectral residue of
$\tU$, call it $\trho_n$ is given by
$$\trho_n=-\frac{2}{n} \Res(\tphi_n;\nu=n).$$
Since 
$$\tphi_n = (n-\nu/2)\phi_n + \text{ lower order terms},$$
it follows
that
$$\trho_n = -\frac{2}{n} \lp n - \frac{n}{2}\rp \rho_n = -\rho_n,$$
where $\rho_n$ is the $n\sth$ spectral residue of $U$.  The theorem is
proved.

Finally, let us see how the relation between the spectral residues and
the prepotential leads us to the inversion formulas \eqref{eq:rhotow}
and \eqref{eq:winv}. With Theorem \ref{th:rhoinv} at our disposal it
is not difficult to describe $W$ in terms of the $\rho_n$.  
\begin{proposition}
  $$W_n=\!\!\sum_{\subover{|\cmp|=n}{\ell(\cmp) \text{\rm\ odd}}}
  \frac{1} {(p_1+p_2)(p_2+p_3)\ldots (p_{l-1}+p_l)}\, \rho_\cmp.$$
\end{proposition}
\begin{proof}
  Note that $2 z W_z = U - \tU$, or equivalently that $W_n =
  (U_n-\tU_n/(2n)$.  Furthermore, Theorems \ref{th:rhoinv} and
  \ref{th:specresdarboux} imply that
  \begin{align*}
    U_n &= \sum_{|\cmp|=n}\frac{(-1)^{l-1}\, n
    }{(p_1+p_2)(p_2+p_3)\ldots (p_{l-1}+p_l)}\,\rho_\cmp  \\
   \tU_n &= \sum_{|\cmp|=n} \frac{-  n
     }{(p_1+p_2)(p_2+p_3)\ldots (p_{l-1}+p_l)}\,\rho_\cmp   
  \end{align*}
  The desired conclusion now follows immediately.
\end{proof}

The inverse relation, i.e. a formula for the spectral residues in
terms of the $W_n$ is shown in equation \eqref{eq:winv}.  Let us prove
it now.  Let $\phi$ and $\tphi$ denote, respectively the
monic solutions of $\cH(\phi)=0$ and $\tH(\tphi)=0$.  Alternatively, we
could consider $\phi$ and $\tphi$ together as the solutions of the
following coupled first-order equations:
\begin{align*}
  &z\phi_z - \frac{\nu}{2}\,\phi+W\phi  = -\frac{\nu}{2}\,\tphi \\
  &z\tphi_z - \frac{\nu}{2}\,\tphi-W\tphi  = -\frac{\nu}{2}\,\phi \\
  &\phi(0) = \tphi(0) = 1
\end{align*}
Introducing sum and difference variables:
$$\sigma = (\phi+\tphi)/2,\qquad \delta=(\phi-\tphi)/2,$$
we obtain
\begin{align*}
  z\sigma_z &= -W\delta, &\sigma(0) &= 1,  \\
  z\delta_z - \nu\delta &= -W\sigma, &\delta(0)&=0
\end{align*}
In terms of power series coefficients we have
\begin{align*}
  n\sigma_n &= - W_{n-1}\delta_1 - \ldots - W_1 \delta_{n-1} \\
  (n-\nu)\delta_n &= - W_n - W_{n-1}\sigma_1 - \ldots - W_1 \sigma_{n-1}
\end{align*}
Now 
$$\Res(\phi_n;\nu=n)=\rho_n,$$
and by Theorem
\ref{th:specresdarboux}, 
$$\Res(\tphi_n;\nu=n)=-\rho_n.$$
Consequently,
$$\rho_n = \Res(\delta_n;\nu=n).$$ 

It is evident that given $W_1, W_2, \ldots$ the above relations can be
solved for $\sigma_n$ and It is also evident that each $\sigma_n$ will
be a sum of products of even numbers of $W_n$ and that each $\delta_n$
will be a sum of products of odd numbers of $W_n$.  Indeed, solving
the above relations for $\delta_n$ we obtain
$$
\delta_n = \sum_{\stackrel{|\cmp|=n}{\scriptscriptstyle \ell(\cmp)\; \text{\rm
      odd}}}
\frac{W_l}{\nu-n}\times \ldots \times \frac{W_{p_4}}{(-s_4)}\times
\frac{W_{p_3}}{\nu-s_3} \times \frac{W_{p_2}}{(-s_2)}\times
\frac{W_{p_1}}{\nu-s_1},
$$
where as usual $\cmp=(p_1,\ldots,p_l)$ is a composition of $n$, and
$s_j=p_1+\ldots+ p_j$.  Taking the residue of the right hand side at
$\nu=n$, equation \eqref{eq:winv} follows immediately.

\section{Applications and Conclusion}
The above discussion indicates how the notion of a spectral residue
can be used to give a formal series treatment of differential
equations with spectral parameter.  Our focus has been on
second-order, one-dimensional problems. This is a well-understood
theory with a large selection of techniques for exact solutions.  It
has been our aim to demonstrate that such techniques lead to useful
information once we make the jump to the formal domain.  The author
feels that the application of the spectral residue methodology to more
general classes of eigenvalue problems, e.g. non-linear and
higher-dimensional linear problems, will yield additional
combinatorial results.

In conclusion, let us also indicate some applications for the
identities and the inversion formulas developed here.  We will
consider two topics: enumeration and integrable systems.  An
application to shape-invariant potentials has been reported on in
\cite{Milson}.  The author hopes that the existence of such
applications will serve as an incentive for a further investigation of
spectral residues.

\subsection{Enumeration problems}
Consider a selection of $k$ triples of numbers $(a,b,c)$ with $1\leq
a<b<c\leq n$. Alternatively we could talk about a $3\times k$
rectangular tableaux whose entries are numbers from $1$ to $n$
(repetitions permitted) with the constraint that the columns be
arranged in ascending order. A natural question in this setting is the
following. If the $k$ triples are selected in a uniformly random
manner, what is more likely: a multiple occurrence of a number in the
second row, or a multiple occurrence in the third row.  The answer is
that duplication is more probable in the third row (or indeed in the
first row by consideration of symmetry) than in the second row.

This fact is straightforward consequence of the identity shown in
\eqref{eq:rhonvfac}.  To see why let us remark that $\binom{m-1}{2}$
is the number of triples with $m$ as the third entry.  Therefore the
generating function that counts tableaux with distinct
entries in the third row is
$$
F(x)=\lp 1+ \frac{1\cdot 2}2\, x\rp \lp 1 + \frac{2\cdot 3}{2}\, x\rp
\ldots \lp 1 + \frac{n(n-1)}2\, x\rp.$$
This is more or less the right
hand side of equation \eqref{eq:rhonvfac}.  Next we note that $m(n-m)$
is the number of triples with $m$ as the middle entry.  Therefore the
generating function for the number of tableaux with distinct entries
in the middle row is given by
$$\prod_{m=1}^{n-1} ( 1+m(n-m)x ) = \sum_{|\cmp|=n} 
\frac{n!}{s_\cmp}\,\frac{n!}{s_{\cmp'}}\; x^{n-\ell(\cmp)}
$$
This is closely related to the left hand side of
\eqref{eq:rhonvfac}.  Indeed, rewriting \eqref{eq:rhonvfac} we have
$$F(x) = \sum_{|\cmp|=n}
\lp \prod_i{\frac{p_i}{2^{p_i-1}}} \rp
\frac{n!}{s_\cmp}\,
\frac{n!}{s_{\cmp'}}\;
 x^{n-\ell(\cmp)}$$
The weights $\prod_i p_i/(2^{p_i-1})$ in the above right-hand
side are less than one, and the desired conclusion follows.

The other identities derived in Section \ref{sect:espot} have similar
enumerative interpretations.  The gist of these interpretations is
that various measures that count sparseness of triples relative to
their largest entries correspond to certain other measures that count
sparseness relative to the middle entries.  These developments will be
reported on elsewhere.

\subsection{Integrable systems.}
It has already been stated that spectral residues are meant to furnish
a kind of spectral information for formal differential equations with
a parameter.  It is natural then to enquire for a physical
interpretation of the spectral residue sequence.  We will show that it
is makes sense to regard the spectral residues as a formalized version
of the potential's scattering data.

Such an interpretation comes about when we inquire about the evolution
of the spectral residues concomitant to the evolution of the
potential function with respect to the KdV equation.
\begin{proposition}
  Suppose that the potential series $U(z,t)$ evolves according to the
  non-commutative version of the KdV equation (see \cite{OlvSok} for a
  list of references):
  $$ U_t = D^3(U) - 3 D(U) U - 3\, U\!D(U),$$
  where $D=z\partial_z$.  Then, the spectral residues $\rho_n$ evolve
  according to the rule
  $$\partial_t(\rho_n) = n^3\rho_n.$$
\end{proposition}
The proof is straightforward; it relies on the following identity
{\small
\begin{multline}
  \label{eq:cubeid}
  (p_1+\ldots +p_n)^3 - (p_1^{\,3} + \ldots + p_n^{\,3}) = \\
\sum_j (p_1+\ldots +
  p_j)(p_{j+1}+\ldots + p_n) (p_j+p_{j+1}).
\end{multline}
}
Thus if we interpret the $n$ in $\rho_n$ as square roots of the energy
( see the discussion following equation \eqref{eq:fullsa} )
we see that the spectral residues evolve according to the same rule as
the scattering data (reflection and normalization coefficients) of 
KdV wave profiles.

We now have the following quite natural interpretation of the
transformations relating the coefficients of $U(z)$ and the spectral
residues, i.e. equations \eqref{eq:rhodef} and \eqref{eq:rhoinv}:
theses inversion formulas describes a formalized version of the
well-known inverse scattering transformation for the KdV equation
\cite{GGKM}.  As such, the soliton solutions correspond to finitely
supported residue sequences, and indeed it is possible to recover the
well known n-soliton formula \cite{Hirota,WadSaw} by means of equation
\eqref{eq:rhoinv}. 

It must be noted that the formula \eqref{eq:rhoinv} and its relation
to solitons has been established previously by Wadati and Sawada
\cite{WadSaw}.  The identity \eqref{eq:cubeid} also occurs in that
paper.  Indeed, these authors went so far as to derive the
Gelfand-Levitan-Marchenko equation from \eqref{eq:rhoinv}.  This is
not surprising, in light of the above interpretation of
\eqref{eq:rhoinv} as the restorative half of the inverse scattering
transform.  The missing idea in Wadati and Suwada's article was
equation \eqref{eq:rhodef}: the forward portion of the formalized
I.S.T.

The next logical step in this development is the representation of
other integrable systems by means of combinatorial inversions.  It is
reasonable to expect that the spectral residue formalism will be
useful in translating the spectral problems that represent integrable
equations into appropriate analogues of \eqref{eq:rhodef}.  The
possible benefits of such a program are two-fold.  First, the
translation of scattering transforms for other integrable systems may
yield new combinatorial results on par with the results for the
Schr\"odinger equation reported on in this article.  Second, by
analogy with harmonic analysis it may be feasible to characterize
classes of solutions to KdV by means of spectral residues.  We already
know that solitons correspond to finitely supported residue sequences.
Is it possible to characterize compactly supported or exponentially
decaying solutions in terms of an appropriately rapid decay of the
residue sequence?  If such an approach is successful, it seems
reasonable that it could be carried over to other integrable
equations.  These matters call for further investigation.

\end{document}